# Soliton emission in amplifying optical lattice surfaces

Yaroslav V. Kartashov, Victor A. Vysloukh*, and Lluis Torner

*ICFO-Institut de Ciencies Fotoniques, and Universitat Politecnica de Catalunya,*

*Mediterranean Technology Park, 08860 Castelldefels (Barcelona), Spain*

We address surface solitons supported by the interface of optical lattices imprinted in saturable media with surface-localized gain. The nonlinearity saturation puts restrictions on the maximal energy flow carried by surface solitons. As a consequence, the presence of thin amplifying layer near the surface results in the controllable emission of solitons towards the lattice at angles depending on the amplification rate and on lattice depth.

Self-action of light at the interface of different nonlinear materials may give rise to formation of surface solitons [1,2]. The experimental observation of such states became possible after their prediction at the interface with semi-infinite waveguide arrays [3], in particular, imprinted in semiconductors with focusing nonlinearity [4]. The interface of optical lattice imprinted in defocusing medium can support gap surface solitons analyzed in [5,6] and observed in [7,8] at the edge of $LiNbO_3$ waveguide arrays. Two-dimensional lattice interfaces also support surface solitons [9-13]. All this investigations have been conducted in passive media. Nevertheless, embedding rare earth ions into such materials as $LiNbO_3$, photorefractive glass waveguides, and other crystals allows combining active and nonlinear properties in a single medium [14,15]. Such photorefractive materials allow technological fabrication [14,15,7,8] or optical induction [16] of lattice interfaces and offer an opportunity to study the impact of both nonlinearity saturation and amplification on surface soliton properties. Note that amplification levels as high as 13 dB in a 7-cm long crystal pumped with some 200 mW have been reported [14]. Substantial amplification can be achieved in Nd-doped photorefractive SBN crystals [15].

In this Letter we reveal that combination of nonlinearity saturation and amplification in a thin layer localized near the lattice surface results in the controllable emission of solitons towards the lattice akin to soliton emission encountered in some conservative settings [17,18]. We show that the emission angle can be controlled by the lattice depth and the amplification rate.

We consider the propagation of a laser beam along the $\xi$-axis near the interface of a semi-infinite optical lattice imprinted in saturable medium in the presence of amplification in a thin layer near the lattice surface. Light propagation is described by the nonlinear Schrödinger equation for dimensionless field amplitude $q$:

$$i\frac{\partial q}{\partial \xi} = -\frac{1}{2}\frac{\partial^2 q}{\partial \eta^2} + \frac{\sigma q|q|^2}{1+S|q|^2} - pR(\eta)q + i\gamma G(\eta)q, \qquad (1)$$

Here $\eta, \xi$ stand for the transverse and longitudinal coordinates normalized to the characteristic transverse scale $x_0$ and diffraction length $kx_0^2$, respectively; $\sigma = \pm 1$ for defocusing/focusing media; the parameter $p$ describes the depth of the optical lattice, defined as $R = 0$ for $\eta < 0$ and $R = [1-\cos(\Omega\eta)]/2$ for $\eta \geq 0$; $S$ is the saturation parameter; $\gamma$ is the amplification coefficient. We assume that amplification occurs in a thin near-surface layer, so that $G = \exp[-(\eta-\eta_\gamma)^2/w_\gamma^2]$, where $\eta_\gamma$ and $w_\gamma$ are the position of center and width of amplification domain. We set $\Omega = 2$ and assume that $w_\gamma = \pi/2\Omega$ is comparable with the lattice period. We suppose that doping with active ions, as well as optical pumping, do not substantially modify the refractive index and do not lead to distortion of lattice refractive-index profile [14,15]. In the particular case of lattices imprinted in SBN crystals biased with a dc electric field $\sim 5$ kV/cm and a laser beam with width $5~\mu$m at $\lambda \approx 1~\mu$m, a length $\xi \sim 1$ corresponds to some $0.38$ mm, $\Omega = 2$ sets lattice period $\sim 16~\mu$m, the parameter $p = 1$ corresponds to a refractive index variation $\sim 4\times 10^{-4}$, and $q = 1$ corresponds to a peak intensity of the order of $100$ mW/cm$^2$.

To understand the mechanism behind soliton emission by the amplifying interface, we first consider stationary solitons at $\gamma = 0$. We search for them numerically in the form $q(\eta,\xi) = w(\eta)\exp(ib\xi)$. Focusing lattice interface supports simplest odd (centered in the first channel) and even (centered between the first and second channels) solitons. At low energy flows $U = \int_{-\infty}^{\infty}|q|^2 d\eta$ (conserved at $\gamma = 0$), both odd and even solitons strongly expand into the lattice and gradually approach each other, which results in appearance of lower cutoff (on $b$) for soliton existence (Fig. 1(a)). Such solitons exist only above a minimal energy flow. At high energy flows, in the strong saturation regime odd and even solitons also approach each other because humps of even soliton gradually become asymmetric (Fig. 1(b)). Thus, nonlinearity saturation results in the appearance of an upper cutoff $b_{\mathrm{upp}}$ for soliton existence. This is because the mean refractive index inside the lattice is higher than that at $\eta < 0$ and soliton in the regime of strong saturation tend to shift deeper into the lattice region rather than into the uniform medium. For high enough $U$ no solitons are located in the first lattice channel, i.e. the saturable interface "repels" high-energy solitons. The $U(b)$ curves for odd and even solitons form closed loops (Fig. 2(a)). Odd solitons are stable in most of their existence domain where $dU/db > 0$, while even solitons are always unstable. The domain of soliton existence becomes narrower with increasing $S$ and shrinks completely at $S = S_{\mathrm{cr}} \approx 3.44$ (Fig. 2(b)). A similar picture takes place for gap solitons at defocusing interfaces, with the only difference that odd gap soliton [5] transforms into twisted gap soliton in the low-energy (Fig. 1(c)) and high-energy (Fig. 1(d)) cutoffs. Such a transformation is accompanied by equalization of the peaks of odd soliton in the first and second lattice channels at high energies, and strong soliton expansion into the lattice at low energies. The energy flow of surface gap solitons can not exceed a certain maximal value (Fig. 2(c)), while dependencies $U(b)$ for odd and twisted gap solitons form closed loops. The cutoffs fall into the first finite gap of the lattice spectrum. They vary with $S$, so that for $S > S_{\mathrm{cr}} \approx 3.44$ the interface can not support solitons from first finite gap residing in the first channel (Fig. 2(d)). Thus, the nonlinearity saturation puts important restrictions on the energy flows carried by the surface solitons.

Once adiabatic ($\gamma \ll 1$) amplification in the vicinity of the first channel is included, the surface solitons adjust their profile to adapt to the local value of $U$ increasing with distance. At a certain propagation length this value exceeds the maximal energy flow that stationary surface soliton could have. This causes the emission of a soliton from the interface towards the lattice (Fig. 3(a)). The radiative losses accompanying the motion of the emitted soliton are remarkably small because of the considerable nonlinearity saturation. Still, increasing the lattice depth reduces the mobility of the emitted solitons and causes a higher radiation rates of the moving solitons. When such radiation is also amplified complex patterns may appear in the vicinity of the amplifying channel. The energy flow of the emitted soliton slightly exceeds the maximal energy of stationary surface solitons. Light that remains in the first channel after soliton emission acts as a seed for the emission of two (Fig. 3(b)), three (Fig. 3(c)), or even extended trains of solitons. The propagation angles and energy flows of all solitons are almost identical to that of the first emitted soliton. Therefore, the thin near-surface amplifying layer acts as a *surface soliton emitter.* Note that such emission does not occur in pure Kerr media where solitons always stay in the near-surface channel, while their amplitude increases and their width decreases upon adiabatic amplification. Soliton emission occurs also in defocusing media. Gap surface solitons require a minimal threshold lattice depth for their existence [5] and it is hard to achieve their emission if the center of amplification channel coincides with the center of first lattice channel. However, when the former is shifted towards the uniform medium $(\eta_\gamma = 0)$, gap solitons acquire additional phase tilt upon amplification facilitating their emission and motion inside the lattice (Fig. 3(d)).

The distance $\xi_e$ at which the first soliton is emitted from the lattice surface in focusing media (we define it as the distance at which the integral soliton center reaches the point $\eta_{int} = 3\pi/\Omega$, i.e. the soliton shifts into the lattice depth by one lattice period) is a monotonically decreasing function of amplification coefficient $\gamma$ (see Fig. 4(a), particularly the inset showing the variation of energy in the first lattice channel with distance). Next we elucidate the dependence of $\xi_e$ on control parameters, the input beam was set to be an exact soliton for $\gamma = 0$. Our simulations revealed that increasing the energy flow of the input light beams reduces the distance $\xi_e$ drastically, but that it does not affect the emission angle $\alpha_e$ (defined as $d\eta_{int}/d\xi$ at $\xi \to \infty$). Thus, at $p = 1$ and $\gamma = 0.04$ the emission distance $\xi_e$ decreases from 113.4 for input energy $U = 2$ to 39.5 for $U = 80$. The angle $\alpha_e$ was found to grow almost linearly with $\gamma$ for high enough amplification coefficients (Fig. 4(b)), while for small $\gamma$ there are oscillations on the $\alpha_e(\gamma)$ dependence that may be connected with small oscillations that the soliton performs inside the first lattice channel before its energy flow becomes high enough for soliton emission.

The emission distance and escape angle can be controlled by acting on the optical lattice depth. Both the emission distance (Fig. 4(c)) and the emission angle (Fig. 4(d)) turn out to be nonmonotonic functions of the lattice depth. The fastest emission occurs for intermediate value of the lattice depth. An initial rapid growth of $\alpha_e$ with $p$ is replaced by a slow decrease for deep lattices where radiative losses become considerable. The qualitatively similar phenomena were found for gap solitons at defocusing interfaces.

Summarizing, we have revealed that a thin amplifying layer located near the edge of a finite optical lattice imprinted in saturable nonlinear media introduces rich soliton emission phenomena. Importantly, the crystal length at which soliton emission occurs, the actual emission angle, and the number of emitted solitons, can be controlled by varying the amplification rate and the optical lattice depth.

*Visiting from the Universidad de las Americas, Puebla, Mexico.

# References with titles

# References without titles

# Figure captions

Figure 1 (color online). Profiles of odd (black curve) and even (red dashed curve) solitons at focusing lattice interface for $b=0.65$ (a), $b=1.4$ (b) and $p=1$, $S=1$. Profiles of odd (black curve) and twisted (red dashed curve) gap solitons at defocusing lattice interface for $b=0.98$ (c), $b=0.2$ (d) and $p=2$, $S=1$.

Figure 2 (color online). (a) Energy flow versus propagation constant for odd (black curve) and even (red dashed curve) surface solitons at $p=1$, $S=1$. (b) Domain of existence of odd and even solitons on the $(S,b)$ plane at $p=1.5$. Panels (a) and (b) correspond to focusing medium. (c) Energy flow versus propagation constant for odd (black curve) and twisted (red dashed curve) gap surface solitons at $p=2$, $S=1$. (d) Domain of existence of odd and twisted solitons on the $(S,b)$ plane at $p=2$. Panels (c) and (d) correspond to defocusing medium.

Figure 3 (color online). Emission of single solitons at $\gamma=0.031$ (a), two solitons at $\gamma=0.04$ (b), and three solitons at $\gamma=0.05$ (c). Panels (a)-(c) correspond to focusing media, $p=1$, $\eta_\gamma=\pi/\Omega$, and input energy flow $U=20$. (d) Emission of gap solitons in defocusing media at $\gamma=0.12$, $p=1.8$, $\eta_\gamma=0$, and $U=60$. White dashed lines indicate interface position. In all cases $S=1$.

Figure 4. Escape distance (a) and angle (b) versus amplification coefficient at $p=0.5$. Inset in (a) shows energy concentrated in the first lattice channel versus $\xi$ for $\gamma=0.01$ (red dashed curve) and 0.02 (black curve). Escape distance (c) and angle (d) versus lattice depth at $\gamma=0.04$. In all cases $S=1$ and energy flow of input soliton $U=17.2$. Focusing medium.

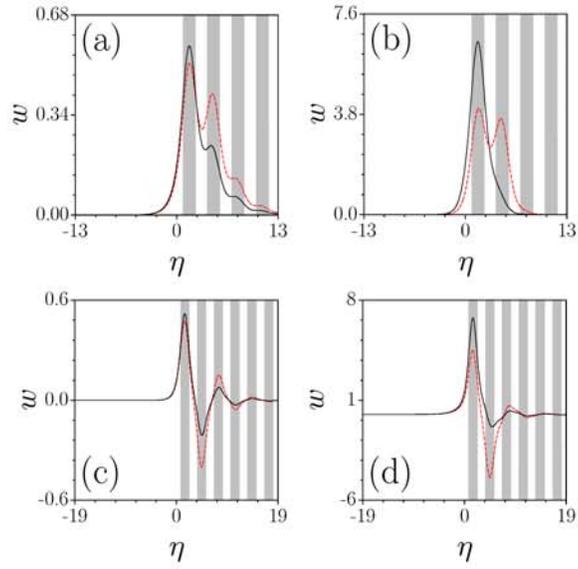

Figure 1 (color online). Profiles of odd (black curve) and even (red dashed curve) solitons at focusing lattice interface for $b = 0.65$ (a), $b = 1.4$ (b) and $p = 1$, $S = 1$. Profiles of odd (black curve) and twisted (red dashed curve) gap solitons at defocusing lattice interface for $b = 0.98$ (c), $b = 0.2$ (d) and $p = 2$, $S = 1$.

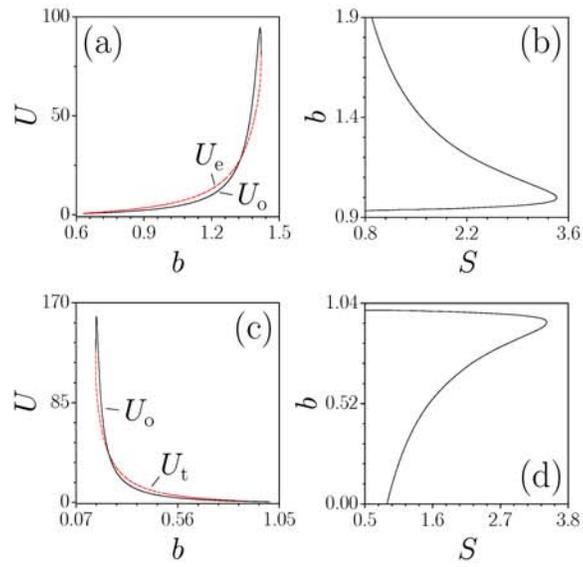

Figure 2 (color online). (a) Energy flow versus propagation constant for odd (black curve) and even (red dashed curve) surface solitons at $p=1$, $S=1$. (b) Domain of existence of odd and even solitons on the $(S,b)$ plane at $p=1.5$. Panels (a) and (b) correspond to focusing medium. (c) Energy flow versus propagation constant for odd (black curve) and twisted (red dashed curve) gap surface solitons at $p=2$, $S=1$. (d) Domain of existence of odd and twisted solitons on the $(S,b)$ plane at $p=2$. Panels (c) and (d) correspond to defocusing medium.

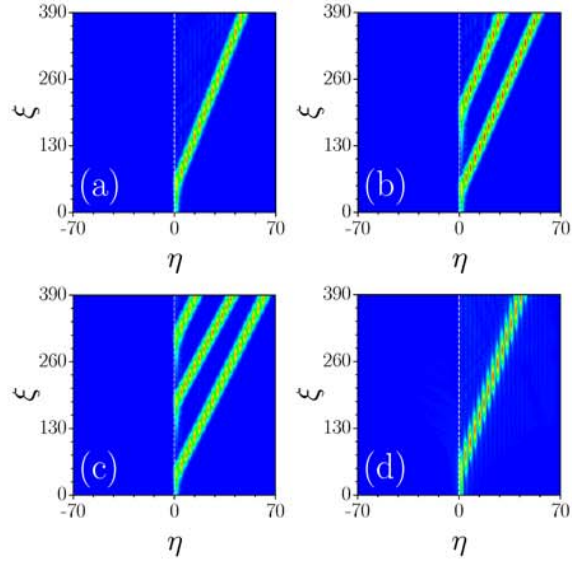

Figure 3 (color online). Generation of single walking soliton at $\gamma = 0.031$ (a), two solitons at $\gamma = 0.04$ (b), and three solitons at $\gamma = 0.05$ (c). Panels (a)-(c) correspond to focusing medium, $p = 1$, $\eta_\gamma = \pi/\Omega$, and input energy flow $U = 20$. (d) Generation of walking gap soliton in defocusing medium at $\gamma = 0.12$, $p = 1.8$, $\eta_\gamma = 0$, and $U = 60$. White dashed lines indicate interface position. In all cases $S = 1$.

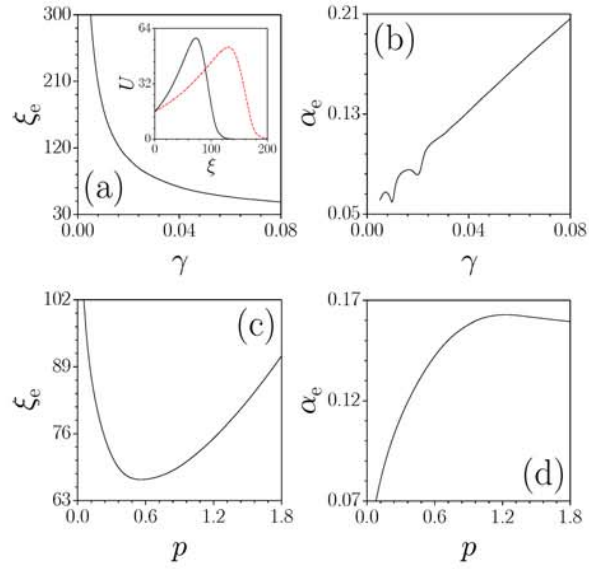

Figure 4. Escape distance (a) and angle (b) versus amplification coefficient at $p=0.5$. Inset in (a) shows energy concentrated in the first lattice channel versus $\xi$ for $\gamma=0.01$ (red dashed curve) and $0.02$ (black curve). Escape distance (c) and angle (d) versus lattice depth at $\gamma=0.04$. In all cases $S=1$ and energy flow of input soliton $U=17.2$. Focusing medium.